\documentstyle[prb,aps]{revtex} 
\tighten
\input epsf
\begin{document}
\draft
\twocolumn[\hsize\textwidth\columnwidth\hsize\csname @twocolumnfalse\endcsname
\title{Cluster growth in far-from-equilibrium particle models with diffusion,
detachment, reattachment and deposition}
\author{F. D. A. Aar\~ao Reis${}^{1,}$\footnote{Email address:
reis@if.uff.br} and R. B. Stinchcombe${}^{2,}$\footnote{E-mail address:
r.stinchcombe1@physics.ox.ac.uk}}
\address{
${}^{1}$ Instituto de F\'\i sica, Universidade Federal Fluminense,
Avenida Litor\^anea s/n, 24210-340 Niter\'oi RJ, Brazil\\
${}^{2}$ Theoretical Physics, Department of Physics, Oxford University, 1 Keble
Road, Oxford OX1 3NP}
\date{\today}
\maketitle
\begin{abstract}
Monolayer cluster growth in far-from-equilibrium systems is investigated by
applying simulation and analytic techniques to minimal hard core particle
(exclusion) models. The first model (I), for post-deposition coarsening
dynamics, contains mechanisms of diffusion, attachment, and slow activated
detachment (at rate $\epsilon\ll 1$) of particles on a line. Simulation shows
three successive regimes of cluster growth: fast attachment of isolated
particles; detachment allowing further ${\left( \epsilon t\right)}^{1/3}$
coarsening of average cluster size; and $t^{-1/2}$ approach to a saturation
size going like $\epsilon^{-1/2}$. Model II generalizes the first one in
having an additional mechanism of particle deposition into cluster gaps,
suppressed for the smallest gaps. This model exhibits early rapid filling,
leading to slowing deposition due to the increasing scarcity of deposition
sites, and then continued power law (${\left( \epsilon t\right)}^{1/2}$) cluster size
coarsening through the redistribution allowed by slow detachment. The basic
${\left( \epsilon t\right)}^{1/3}$ domain growth laws and $\epsilon^{-1/2}$
saturation in model I are explained by a simple scaling picture involving the
time for a particle to detach and diffuse to the next cluster. A second,
fuller approach is presented which employs a mapping of cluster configurations
to a column picture and an approximate factorization of the cluster
configuration probability within the resulting master equation. This allows,
through the steady state solution of the corresponding equation for a cluster
probability generating function, quantitative results for the saturation of
model I in excellent agreement with the simulation results. For model II, it
provides a one-variable scaling function solution for the coarsening
probability distribution, and in particular quantitative agreement with the
cluster length scaling and its amplitude.\\
\end{abstract}
\pacs{PACS numbers: 05.40.-a, 05.50.+q, 68.43.Jk}
\narrowtext
\vskip2pc]

\section{Introduction}

This paper is concerned with domain growth in far from equilibrium systems.
This is a subject of increasing interest both for the wide range of behaviors
and for the large number of applications, which range from phase separation in
mixtures to island formation and coarsening during deposition of a thin film
or submonolayer~\cite{bartelt92,bales}, among other systems.

Our aim is to discuss a series of one-dimensional exclusion models with
particle diffusion, reversible or irreversible attachment to clusters and
deposition mechanisms that represent, for example, volume reduction effects
after cluster coalescence. Diffusion processes tend to bring these systems to
equilibrium steady states, but pressure and other external influences may drive
the system to new steady states. Though not usually related to a specific real
problem, these one-dimensional models may reveal interesting features
that help to understand more complex and realistic surface and bulk
systems~\cite{sollich,jackle}, with the advantage of being more tractable both
analytically and numerically. We will discuss a series of plausible physical
situations in systems with diffusion and mechanisms that drive them out of
equilibrium, in order to understand the details of domain growth and
convergence to steady states, if it occurs.

In the first model, hereafter called model I, a fixed fraction $\rho$ of a
one-dimensional lattice is randomly filled with hard core particles. The
diffusion rates are $r=1$ when they are free, i. e. when they have two
empty nearest-neighbor sites, and $r=\epsilon\sim e^{-E/T}$ (where $E$ is the
related energy barrier) when they have one occupied nearest neighbor site
(Fig. 1a). For $\epsilon\ll 1$ ($T\to 0$), the average aggregates' length
grows as $t^{1/3}$ in a long time range, and eventually approaches saturation
at $\sim {\epsilon}^{-1/2}$ with a slow $t^{-1/2}$ decay (Sec.
\ref{secdiffusiononly}). In the limit $\epsilon\to 0$, this model is equivalent
to the Ising model with Kawasaki dynamics previously studied by Cornell et
al~\cite{cornell}, who focused on its zero temperature features. However,
the dynamic rules are mainly motivated by the Clarke-Vvedensky model for
thin films or submonolayer growth~\cite{cv}, excluding the deposition
processes. In the simplest versions of that model, an isolated adatom has to
overcome an energy barrier $E_s$ to diffuse, while when it is attached to
$n$ nearest neighbors the energy barrier increases to $E_s+nE_b$, where $E_b$
is a bonding energy. This model and related ones were already intensively
studied in two dimensions during the deposition
process~\cite{cv,barkema,ratch}, but a few works have considered the
post-deposition coarsening dynamics~\cite{lam}.

Subsequently we will generalize the previous hard core dynamical system by
introducing deposition of particles (see e. g. Refs. \protect\cite{evans} and
\protect\cite{robinbook}), but allowing deposition only at (empty) sites with at least
one empty nearest neighbor (Fig. 1b). In this model (referred to as model II),
the domain coalescence, which generates larger vacancies between aggregates,
is followed by a density increase. The exclusion of deposition at single holes
between clusters represent the geometrical frustration of real systems. In
Sec. \ref{secdiffusiondep}, we will show that this model exhibits a $t^{1/2}$
domain growth. This is among other results we have obtained by simulation
studies, which are presented for both models I and II in Secs.
\ref{secdiffusiononly} and \ref{secdiffusiondep}, respectively.

\begin{figure}
\epsfxsize=8,5cm
\begin{center}
\leavevmode
\epsffile{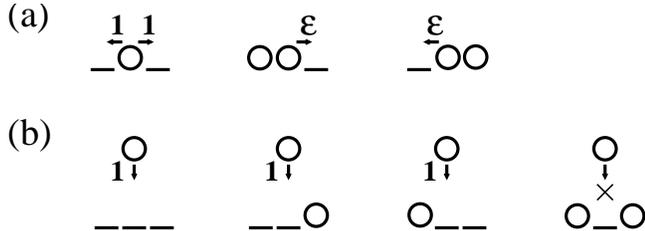}
\caption{(a) Diffusion and detachment processes of model I, with the
corresponding rates. (b) Added deposition processes of model II, with the
corresponding rates. The deposition at sites with two occupied neighbors
(last process) is forbidden.}
\label{fig1}                        
\end{center}
\end{figure}     

Our models share some aspects with diffusion limited coalescence
models~\cite{abad,benavraham,zhong} and with fragmentation-aggregation
models~\cite{suphriaetal,suphriarobin}. They are similarly described in terms
of cluster or interval probabilities, and like the fragmentation models they
are amenable to analytic investigation based on an independent cluster
approximation (the independent interval approximation to the joint cluster
length probability occurring in the master equation). We use this approach to
explain properties of models I and II, including distributions of cluster size
(Secs. \ref{secdiffusiononly} and \ref{secdiffusiondep}). Further, a simple
scaling picture can be developed in order to describe the basic domain growth
laws; we use this at the beginning of Sec. \ref{secdiffusiononly}.

\section{Model I: diffusion, detachment and reattachment of particles}
\label{secdiffusiononly}

\subsection{Processes}
\label{Processdifonly}

The model studied in this section has the particle hopping processes depicted
in Fig. 1a. Isolated particles hop symmetrically on a chain at unit rate
("diffusion"), while a single particle with a left hand / right hand neighbor
can hop to an empty right / left neighbor with rate $\epsilon$ (detachment).
So clusters evolve by detachment and reattachment of particles. The model is
of exclusion type: no site can accommodate more than one particle.

This model is clearly particle-conserving, so density $\rho$ is fixed. The case
$\epsilon\ll 1$ is of particular interest since, as reported in the simulation
studies below and explained in the following subsection, very large clusters
emerge.

\subsection{Simulations}
\label{Simulationsdifonly}

We simulated model I in one-dimensional lattices of length $L=8000$. This
length is sufficiently large to ensure that finite size effects are negligible,
as shown by comparisons of some results with data from lattices with
$L=16000$ (particularly for the smallest values of $\epsilon$ this
comparison is essential).

Initially, the lattice is randomly filled with a density of
particles $\rho$. We simulated three values of the density, $\rho = 0.1$,
$\rho = 0.5$ and $\rho = 0.9$, which are representative of the range of
intermediate densities, i. e. densities not too small ($\rho\approx 0$) nor
too large ($\rho\approx 1$). For $\rho = 0.1$ and $\rho = 0.9$, we considered
several values of the diffusion rate $\epsilon$ ranging from $\epsilon =
{10}^{-1}$ to $\epsilon = {10}^{-3}$, and for $\rho = 0.5$ we performed
simulations until $\epsilon = {10}^{-5}$.

The sequence of characteristic behaviors of model I, as shown by simulation
results, are: $(i)$ early fast attachment of isolated
particles to each other to form clusters; $(ii)$ an intermediate regime in
which detachment sets in, allowing further coarsening; $(iii)$ finally, there
is a diffusive approach to a saturated state where the clusters have a large
steady mean size that depends on $\epsilon$.

The three regimes are well separated at small $\epsilon$. This is illustrated
in the plot of $\log_{10}{d}$ versus time $\log_{10}{t}$, shown in Fig. 2,
where $d$ is the mean size of clusters of two or more particles; $d$ is given
in terms of the probability $P_t(m)$ that an arbitrarily chosen cluster has
size (or mass) $m$ at time $t$ by
\begin{equation}
d = {{\sum_{m=2}^{\infty}{mP_t\left( m\right)}}\over
{\sum_{m=2}^{\infty}{P_t\left( m\right)}}} .
\label{defd}
\end{equation}

\begin{figure}
\epsfxsize=8,5cm
\begin{center}
\leavevmode
\epsffile{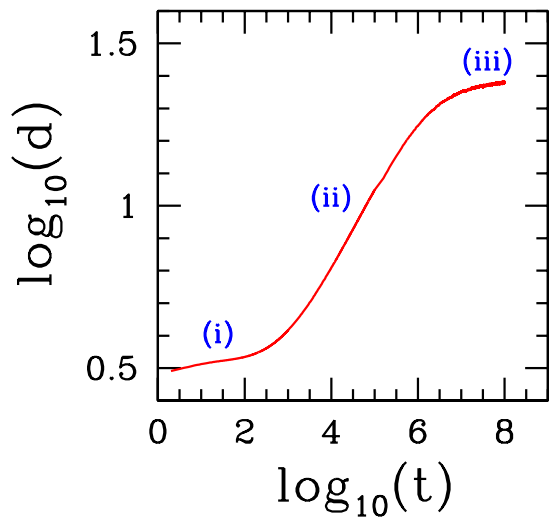}
\caption{Typical time evolution of the average cluster length
$d$ in model I, with three different regimes. Data in the plot were obtained
for $\rho = 0.5$ and $\epsilon = {10}^{-4}$.}
\label{fig2}                        
\end{center}
\end{figure}     

The early time dependence of $d$ in region $(i)$ (at small $\epsilon$) starts
with a characteristic increase with rate proportional to a high power of
$\epsilon$, and then crosses over to a form allowing data collapse in terms of
the reduced time variable $\epsilon t$, as shown in Fig. 3.

\begin{figure}
\epsfxsize=8,5cm
\begin{center}
\leavevmode
\epsffile{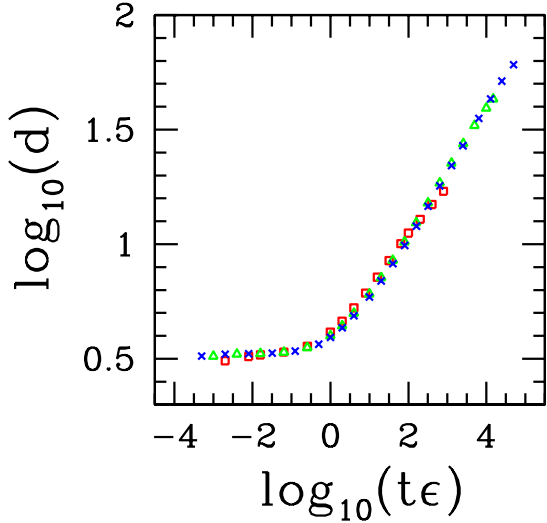}
\caption{Average cluster length as a function of the scaling
variable $\log_{10}{\left( \epsilon t
\right)}$, for model I with $\rho=0.5$ and $\epsilon
={10}^{-3}$ (squares), $\epsilon ={10}^{-4}$ (triangles) and $\epsilon
=5\times {10}^{-5}$ (crosses).}
\label{fig3}                        
\end{center}
\end{figure}     

\begin{figure}
\epsfxsize=8,5cm
\begin{center}
\leavevmode
\epsffile{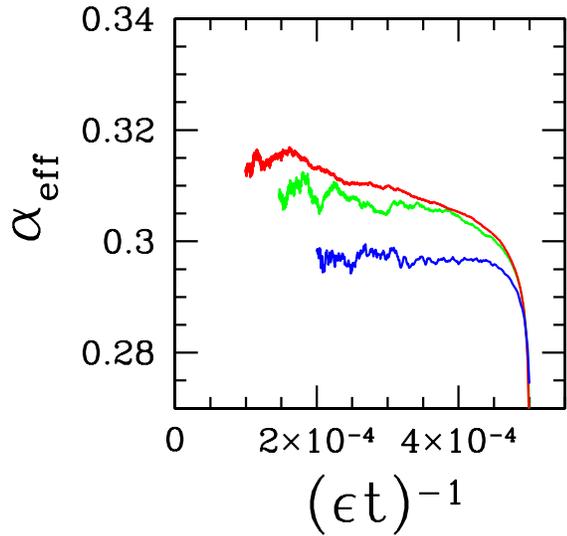}
\caption{Effective exponents $\alpha$, defined as local slopes of the
$\log{d}\times\log{t}$ plots, for model I with $\rho=0.5$ and $\epsilon
=5\times {10}^{-5}$ (below, blue), $\epsilon =2\times{10}^{-5}$ (medium,
green) and $\epsilon ={10}^{-5}$ (up, red).}
\label{fig4}                
\end{center}
\end{figure}     

In region $(ii)$, $d$ increases as
\begin{equation}
d \sim t^{\alpha} .
\label{defalpha}
\end{equation}
The apparent exponent $\alpha_{eff}$, defined as the local slope of the
$\log{d}\times\log{t}$ plot, was calculated numerically. $\alpha_{eff}$ is
shown in Fig. 4 as a function of ${\left( \epsilon t\right)}^{-1}$ for three
different values of $\epsilon$. It appears to approach the value $\alpha = 1/3$
in the limit of small $\epsilon$ and correspondingly large $t$, which
is consistent with the prediction of a simple scaling description (next
subsection).

Fig. 5 shows the diffusive approach of the mean cluster size to its saturation
value $d_\infty$. This approach is well described by
$d = d_\infty - {C\over{t^{1/2}}}$, for $t\to\infty$, with $C$ constant.
The dependence on $\epsilon$ of the saturation value $d_\infty$ is illustrated
in Fig. 6 for $\rho=0.5$. The least squares fit in Fig. 6 gives
\begin{equation}
d_\infty \approx 0.72\epsilon^{-1/2} + 1.93,
\label{scalingdinf}
\end{equation}
in which the dominant (proportional to $\epsilon^{-1/2}$) and the
sub-dominant (additive constant) terms were estimated. Like Eq.
(\ref{defalpha}) with $\alpha = 1/3$, this result follows from the analytic
work in Sec. \ref{Theorydifonly}.

\begin{figure}
\epsfxsize=8,5cm
\begin{center}
\leavevmode
\epsffile{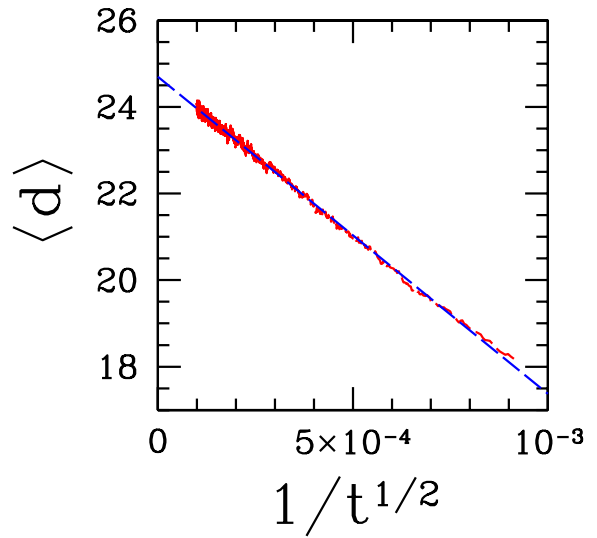}
\caption{Average cluster size at long times $t$ as a function of
$1/t^{1/2}$, for $\rho=0.5$ and $\epsilon ={10}^{-3}$ in model I. The dashed
line is a least squares fit of the data for ${10}^6\leq t\leq {10}^8$.}
\label{fig5}                        
\end{center}
\end{figure}     

\begin{figure}
\epsfxsize=8,5cm
\begin{center}
\leavevmode
\epsffile{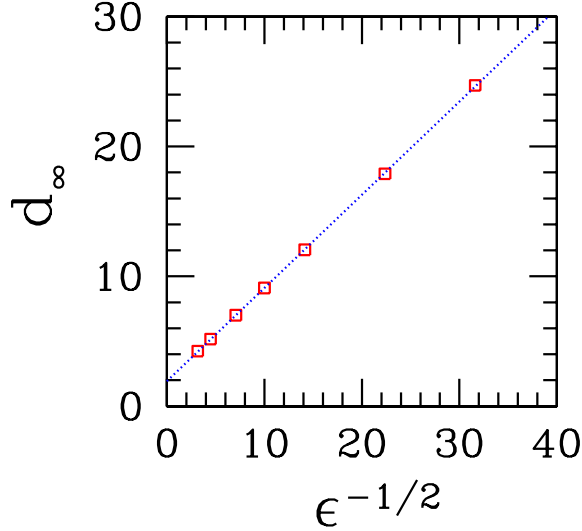}
\caption{Saturation value of the average cluster length as a
function of $\epsilon^{-1/2}$ in model I. The dotted line is a least squares
fit of the data.}
\label{fig6}                
\end{center}
\end{figure}     

\subsection{Theory}
\label{Theorydifonly}

The characteristic results just described have been interpreted by simple
heuristic scaling arguments and by detailed analytic studies starting from the
master equation and employing an independent interval approximation. This
second approach is capable of yielding cluster length distributions and their
evolution.

To begin with, we focus on the asymptotic cluster size at small $\epsilon$.
This asymptotics occurs in the regime where lone particles are rare, and those
that are present are in the process of reattaching themselves to the clusters
they came from or to a neighboring one. The second case provides the
sharing which sets the mean cluster size $d$. At densities of order $\rho\sim
1/2$, the cluster size is roughly of the order of cluster separation (see Fig.
7a). Thus the equilibrium of detachment time ($1/\epsilon$) and time of
diffusion to a neighboring cluster ($\sim d^2$) gives the observed saturation
result
\begin{equation}
d \sim \epsilon^{-1/2} .
\label{scalingdinf2}
\end{equation}
This argument can be generalized rather obviously to explain the $t^{-1/2}$
approach to saturation.

A more interesting application is the
explanation of the early cluster size growth law (Eq. \ref{defalpha}). Here,
unlike the saturation just described, the cluster separations are such that
the detached particle is likely to return and reattach many times before it
eventually diffuses to the next cluster (Figs. 7a and 7b). Its likelihood of
returning to the origin means that the detachment rate $\epsilon$ needs to be
replaced by an effective rate $\tilde{\epsilon} = \epsilon P_{mig}$, where
$P_{mig}$ is the (migration) probability that a freely diffusing particle does
not return to the detachment site before diffusing the distance $\sim d$ to
the next cluster. In other words (see Fig. 7b), this is the probability that a
free particle at position $y=1$ does not return to the origin ($y=0$) before a
time of order $d^2$, which is the typical time for diffusion along a distance
$d$. Considering that
\begin{equation}
Q(y,t) = {y\over{{\left( 4\pi D\right)}^{1/2}}} t^{-3/2}
\exp{\left( - {{y^2}\over{4Dt}}\right)}
\label{probfirstpassage}
\end{equation}
is the probability that the first passage of a random walker at
point $y$ occurs at time $t$~\cite{montroll} ($D$ is the diffusion
coefficient), $P_{mig}$ is given by
\begin{equation}
P_{mig} \sim \int_{d^2}^{\infty}{Q\left( 1,t\right) dt} \sim d^{-1} .
\label{probmig}
\end{equation}

\begin{figure}
\epsfxsize=8,5cm
\begin{center}
\leavevmode
\epsffile{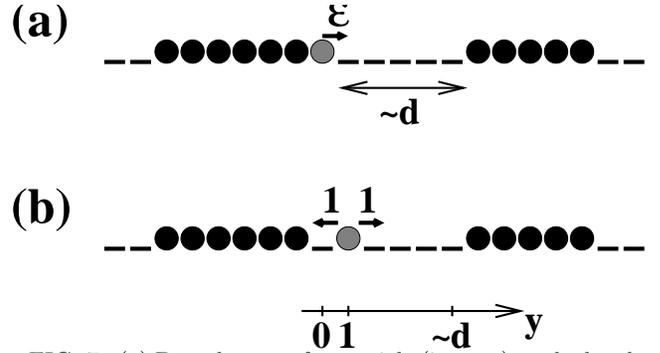}
\caption{(a) Detachment of a particle (in grey) at the border of a cluster,
with rate $\epsilon$. In model I, the mean cluster length is $d$ and the
typical cluster separation is, asymptotically, also of order $d$ for densities
not too small nor too large. (b) A free particle (in grey) immediately after
its detachment from the right cluster.}
\label{fig7}                
\end{center}
\end{figure}     

In terms of the effective rate $\tilde{\epsilon}$, the required time for a
particle to transfer to the next cluster is of order $1/\tilde{\epsilon}$.
Thus, the time required for doubling the size $d$ of a cluster by successive
gain and loss of particles is $d^2/\tilde{\epsilon}\sim d^3/\epsilon$. So the
cluster growth proceeds according to
\begin{equation}
{d\over{dt}}d \sim {d\over{\left( d^3/\epsilon\right)}} ,
\label{eqdif_difonly}
\end{equation}
and hence
\begin{equation}
d\sim {\left(\epsilon t\right)}^{1/3} .
\label{scalingii}
\end{equation}
This explains the behavior seen in the simulations (Figs. 3 and 4). The
situation is analogous to domain scaling in Ising chains where, with Kawasaki
dynamics~\cite{cornell}, spins split off from domain edges and migrate across
to increase the domain size by one lattice unit.

We turn next to the more powerful analysis starting from a version of the
master equation, which can provide a full description of the process. This is
more easily set up by reformulating the process using a column picture, in
which a column of height $m$ represents a cluster of size $m$, and then the
original detachment and diffusion processes correspond to those shown in Fig.
8. Since one cluster has two edges but corresponds to a single column, the
one-particle detachment rate in the column picture is
\begin{equation}
\gamma = 2\epsilon .
\label{rates}
\end{equation}

\begin{figure}
\epsfxsize=8,5cm
\begin{center}
\leavevmode
\epsffile{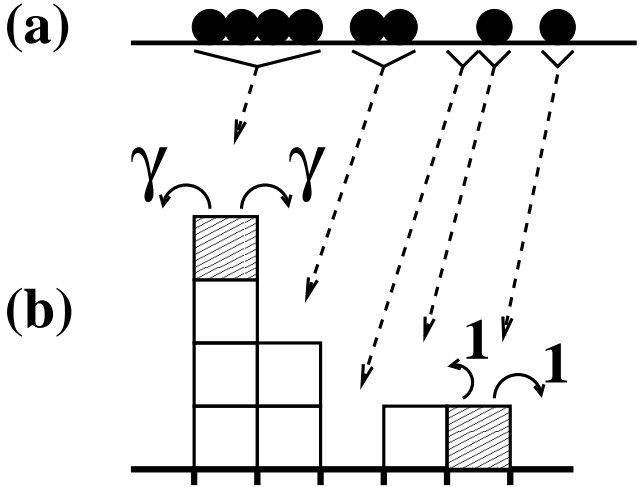}
\caption{(a) Example of particle-hole configuration on a line and the map
(dashed arrows) into a column problem. (b) The processes of particle
detachment from clusters, with rate $\gamma$, and of free particles
diffusion, with rate $1$, in the corresponding column picture.}
\label{fig8}                
\end{center}
\end{figure}     

We denote by $P_t(m)$ the probability that a randomly chosen cluster
(equivalently, column) has size $m$ at time $t$. Then the gain/loss from
in/out processes provides the following master equation, in an independent
interval approximation in which joint probabilities are factorized:
\begin{eqnarray}
&&P_{t+1}(m) - P_t(m) = {\cal A}_m \equiv
\nonumber\\
&&P_t(m-1)\theta (m-1)
\left[ \gamma\sum_{m'\geq 2}{P_t(m')} + 1\times P_t(m'=1) \right] +
\nonumber\\
&&P_t(m+1) \left[ \gamma\theta (m+1-2) + 1\times \delta_{m,0} \right] -
\nonumber\\
&&P_t(m) [ \gamma\theta (m-2) + 1\times \delta_{m,1} +
\nonumber\\
&&\gamma\sum_{m'\geq 2}{P_t(m')} + 1\times P_t(m'=1) ] .
\label{masterdifonly}
\end{eqnarray}
The corresponding equation for the generating function
\begin{equation}
G_t(s) \equiv \sum_{m=0}{P_t(m) s^m}
\label{generating}
\end{equation}
is
\begin{eqnarray}
G_{t+1}(s) &=& G_t(s) \left[ 1 + s a(t) + {\gamma\over s} - \gamma - a(t)
\right]
\nonumber\\
&&+ s(\gamma -1)P_t(1) +
\nonumber\\
&&\left[ \gamma P_t(0) + (1-\gamma)P_t(1) \right] - {\gamma\over s}P_t(0) ,
\label{gendifonly}
\end{eqnarray}
where $a(t) = \gamma \left[ 1-P_t(0) \right] + (1-\gamma) P_t(1)$. It is
easy to check probability and mass conservation using $G_t(0)$ and $G'_t(0)$.

The steady state distribution $P(m)$ and generating function $G(s)$ resulting
from Eq. (\ref{gendifonly}) are given by
\begin{equation}
G(s) = {\left( \gamma - sa\right)}^{-1}
\left[ \gamma P(0) - s(1-\gamma)P(1) \right] ,
\label{gsteadydifonly}
\end{equation}
\begin{equation}
P(m) = {\left( {A\over\gamma}\right)}^{m-1}P(1) , m>1 ,
\label{psteadydifonly}
\end{equation}
and
\begin{equation}
P(1) = A P(0) ,
\label{p1difonly}
\end{equation}
with $A = \gamma \left[ 1-P(0)\right]
{\left[ 1 - (1-\gamma )P(0)\right]}^{-1}$. So the steady state cluster size
distribution is exponential. The mean size of multi-particle clusters (Eq.
\ref{defd}) and the mean mass
$\langle m\rangle \equiv\sum_{m=0}^{\infty}{mP\left( m\right)}$ are then
obtainable in terms of $P(0)$, as is the density $\rho$. So, in particular the
mean cluster length $d\equiv \langle m\rangle$ can be found in terms of $\rho$.
The result simplifies at small $\gamma$ (small $\epsilon$) to
\begin{eqnarray}
d &=& \gamma^{-1/2}
{\left[ {\rho\over{\left( 1-\rho\right)}}\right]}^{1/2} +
{{\left( {\rho/{\left( 1-\rho\right)}}+3\right)}\over 2}
=
\nonumber\\
&&\epsilon^{-1/2}
{\left[ {\rho\over{2\left( 1-\rho\right)}}\right]}^{1/2} +
{{\left( {\rho/{\left( 1-\rho\right)}}+3\right)}\over 2} ,
\label{dsteadydifonly}
\end{eqnarray}
where the dominant and the first sub-dominant terms are shown.
This form is consistent with the scaling result (\ref{scalingdinf2}) and
is in very good agreement with the simulation result (\ref{scalingdinf}),
including the sub-dominant constant term: Eq. (\ref{dsteadydifonly}) gives
$d=0.7071\dots\epsilon^{-1/2}+2$ for $\rho=0.5$ (see also Fig. 6). In the same
limit, this is also, apart from a numerical factor, the characteristic size in
the exponential cluster mass distribution.

\section{Model II: diffusion and deposition of particles}
\label{secdiffusiondep}

\subsection{Processes}
\label{Processdepdif}

Model II is a generalization of model I, different only by having the
deposition processes depicted in Fig. 1b, in addition to the diffusion and
detachment processes of Fig. 1a. This makes the model
non-particle-conserving, which leads to continued coarsening and other scaling
properties and crossover.

\subsection{Simulations}
\label{Simulationsdepdif}

The characteristic behavior of model II, as exhibited by simulation results,
is as follows. For initial densities not too near $\rho =1$, there is: (i) an
early regime of rapid filling, due to deposition, and cluster evolution due to
both processes; (ii) an intermediate regime where deposition slows because of
the scarcity of deposition sites due to the increased density - the exclusion
constraint of course applies. The slow detachment process allows
redistribution of particles, opening up new deposition sites and allowing the
continually slowing coarsening (with no saturation as $\rho <1$).

Fig. 9 shows simulation results for the evolution of the mean cluster size
$d$. That plot shows that $d(t)$ is well fitted by the form
\begin{equation}
d(t) = B t^{1/2} \left( 1 + Ct^{-1/2} + \dots \right) .
\label{ddiffusiondep}
\end{equation}
In Fig. 10 we show the ratio between the estimates of the amplitude $B$ and
$\epsilon^{1/2}$ for several values of $\epsilon$. Those results
give
\begin{equation}
B(\epsilon) \sim b \epsilon^{1/2} ,
\label{scalingB}
\end{equation}
with negligible corrections to scaling, where $b=0.252\pm 0.002$.
This result is in accord with theoretical analysis given in the next
subsection, including the estimate of the amplitude $b$.

\begin{figure}
\epsfxsize=8,5cm
\begin{center}
\leavevmode
\epsffile{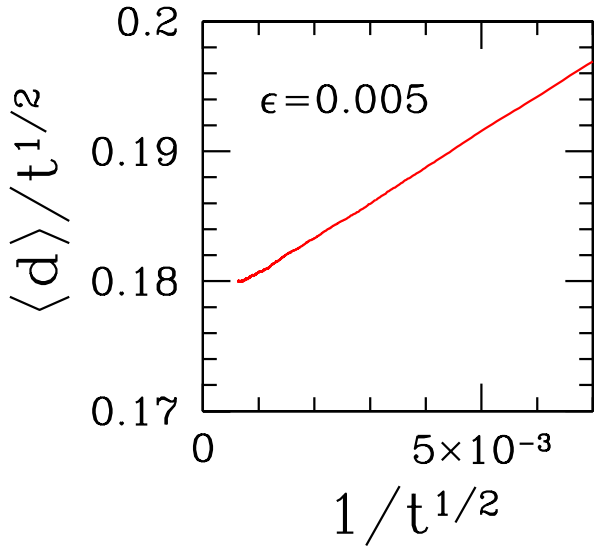}
\caption{Long time scaling of the mean cluster length $\langle
d\rangle$ in model II.}
\label{fig9}                
\end{center}
\end{figure}     

The dependence of the evolving density on time $t$ and rate $\epsilon$ has
also been studied. The simulation results shown in Fig. 11 imply that the
density is a function of the scaling variable ${\left( \epsilon t\right)
}^{1/2}$ and, at very long times, it converges to $1$ as
\begin{equation}
1-\rho \sim {\left( \epsilon t\right) }^{-1/2} .
\label{scalingrho}
\end{equation}
This form is also in agreement with the theoretical analysis below.

\begin{figure}
\epsfxsize=8,5cm
\begin{center}
\leavevmode
\epsffile{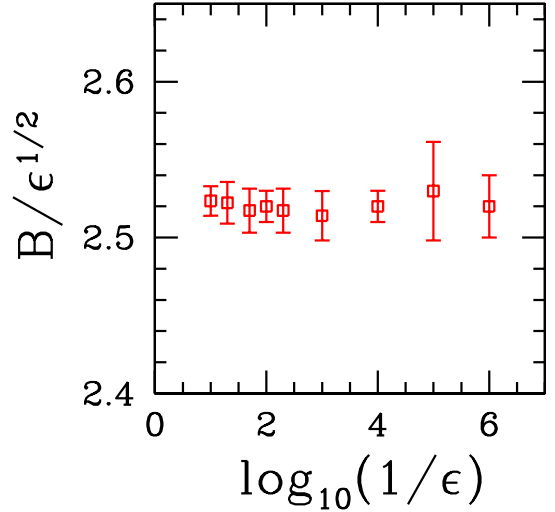}
\caption{Ratio between the estimated amplitude $B$ of the
scaling of average cluster length (Eq. \ref{ddiffusiondep}) and
$\epsilon^{1/2}$, as a function of $\log_{10}{\left( 1/\epsilon\right)}$, in
model II.}
\label{fig10}                
\end{center}
\end{figure}     

\begin{figure}
\epsfxsize=8,5cm
\begin{center}
\leavevmode
\epsffile{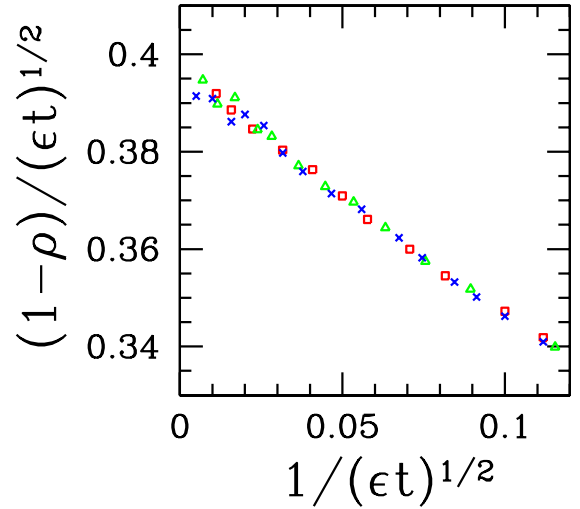}
\caption{Scaling plot of the particle density $\rho$ in model
II, for $\epsilon ={10}^{-2}$ (squares), $\epsilon =5\times {10}^{-3}$
(triangles) and $\epsilon ={10}^{-3}$ (crosses).}
\label{fig11}                
\end{center}
\end{figure}     

\subsection{Theory}
\label{Theorydepdif}

The characteristics presented in the foregoing subsection can be interpreted
using an analytic investigation along the lines of the detailed discussion
given in Sec. \ref{Theorydifonly}.

We have to include the effects of the extra
deposition process, which leads to the decrease of the total number of
clusters and of the number of holes between the clusters as time increases.
On the other hand, the length $L$ of the line in which particles are
deposited and diffuse is kept constant. Consequently, in order to adopt
the column picture of Sec. II.3 (see Fig. 8), it is necessary to consider that
the length $L_0$ of the corresponding column problem decreases in time
(these lengths are related as $L_0=L-M$, where $M$ is the total mass or
total number of particles, for periodic boundary conditions).

The evolution equation here is written for cluster numbers as
\begin{equation}
N(m,t+1) - N(m,t) = L_0 \left( {\cal A}_m + {\cal B}_m \right) ,
\label{masterdiffusiondep}
\end{equation}
where the diffusion contribution ${\cal A}_m$ is given in Eq.
(\ref{masterdifonly}) and the deposition contribution is
\begin{eqnarray}
&&{\cal B}_m = P_t(0) [ 2 \theta\left( m-2\right) P_t\left( m-1\right) -
\nonumber\\
&&2 \theta\left( m-1\right) P_t\left( m\right) + \delta_{m,1}P_t\left( 0\right)
- 2\delta_{m,0} ] .
\label{contribdep}
\end{eqnarray}
The length of the lattice in which the column problem is defined varies due to
deposition as
\begin{equation}
{{L_0(t+1)-L_0(t)}\over{L_0(t)}} = -P_t(0) \left[ 2-P_t(0)\right] .
\label{lenght}
\end{equation}
In these equations, the cluster probability is
\begin{equation}
P_t(m) = {{N(m,t)}\over{L_0(t)}} .
\label{probdifdep}
\end{equation}
They preserve conservation of probability, but mass is no longer conserved.

The resulting equation is similar to ones occurring in coalescence
models~\cite{abad,benavraham,zhong}. From this we expect that a large time and
small $\epsilon$ limit discussed subsequently is equivalent to the model in
Ref. \protect\cite{abad}. Our approach, which exploits the generating function
method, becomes equivalent, in the scaling limit, though in a conjugate space,
to continuum approximations used in the coalescence studies of Refs.
\protect\cite{abad} and \protect\cite{zhong}.

Now the generating function (Eq. \ref{generating}) satisfies
\begin{eqnarray}
&&G_{t+1}(s) {\left[ 1-P_t(0)\right]}^2 - G_t(s) =
\nonumber\\
&&(s-1) \left[ G_t(s) \left( a(t)-{\gamma\over s}\right) + (\gamma -1)P_t(1) +
{\gamma\over s}P_t(0) \right] +
\nonumber\\
&&P_t(0) \left[ 2(s-1)G_t(s) - sP_t(0) + 2\left( P_t(0)-1\right) \right] .
\label{gendifdep}
\end{eqnarray}
In the right hand site of Eq. (\ref{gendifdep}), the first term corresponds to
diffusion processes and the second one to deposition processes.

Because deposition slowly fills the system, we expect the configurations to
coarsen and presumably to go into some scaling asymptotics where mass scales
with some power of $t$, and $P_t(m)$ and $G_t(s)$ each become one-variable
scaling functions. So we look for a long time scaling solution of the above
equation.

At long times, the finite difference $G_{t+1}(s)-G_t(s)$ in Eq.
(\ref{gendifdep}) can be taken as a derivative. The scaling variable will be
some combination of $t$ (large) and $u\equiv 1-s$ (small), the latter because
large cluster sizes arise from structure in $G_t(s)$ at $s\approx 1$. The
variable $u$ is actually conjugate to $m$ (see below). Coarsening will
correspond to the scale of $m$ as $t^z$, with some power $z$, in which case
the one-variable form will be
\begin{equation}
G_t(s) = u^\alpha f\left( ut^z\right) ,
\label{scalinggendifdep}
\end{equation}
with some function $f$.
Normalization requires $\alpha = 0$ and $f(0) = 1$. In the scaling limit, the
relationship of the generating function to the probability $P_t(m)$ requires
the latter to be of the form
\begin{equation}
P_t(m) = {1\over{t^z}} g\left( {m\over {t^z}}\right) ,
\label{scalingpdifdep}
\end{equation}
with
\begin{equation}
f(x) = \int_0^\infty{ g(y) e^{-xy} dy} .
\label{deff}
\end{equation}
It turns out that the consistent scaling solution has $g(0)=0$, so the
$1/t^z$ contribution to $P_t(0)$ vanishes, leaving a leading term of
lower-than-scaling order,
\begin{equation}
P_t(0) = {{c/2}\over{t^{2z}}} ,
\label{Pt0depdiff}
\end{equation}
where $c$ is a constant. Eq. (\ref{gendifdep}) leads to the dynamical
exponent
\begin{equation}
z=1/2
\label{zdepdiff}
\end{equation}
and to the following equation for the one-variable scaling function:
\begin{equation}
xf'(x) - 2cf(x) -2\gamma x^2 f(x) + 2c = 0 .
\label{eqdiff}
\end{equation}

Even without solving Eq. (\ref{eqdiff}) we can infer that
\begin{equation}
P_t(0) = {{c/2}\over t} ,
\label{p0difdep}
\end{equation}
\begin{equation}
{\langle m\rangle}_t  \sim \gamma^{1/2}t^{1/2} \sim \epsilon^{1/2}t^{1/2}
\label{mdifdep}
\end{equation}
and
\begin{equation}
1-\rho_t \sim \gamma^{-1/2}t^{-1/2} \sim \epsilon^{-1/2}t^{-1/2} .
\label{rhodifdep}
\end{equation}
These hold in the long time scaling limit we have introduced and agree with
the observed simulation results in Eqs. (\ref{ddiffusiondep}) ,
(\ref{scalingB}) and (\ref{scalingrho}).

Eq. (\ref{eqdiff}) can be formally solved for the scaling
function $f(x)$ by using the variable $\zeta = x^2$ and considering the
function $f(x) x^{-2c}$. The result is
\begin{equation}
f(x) = {c\over\gamma}\int_0^\infty{ {\left( 1+{v\over\gamma}\right)}^{-c-1}
e^{-vx^2}dv } .
\label{f}
\end{equation}
The large $x$ expansion of $f(x)$ and Eq. (\ref{deff}) provides the small $y$
expansion of $g(y)$:
\begin{equation}
g(y) = {1\over{\gamma^{1/2}}} {{\cal G}{\left(
{y\over{\gamma^{1/2}}}\right) }} ,
\label{g}
\end{equation}
where
\begin{equation}
{\cal G}(u) = \sum_{m=0}^\infty{ {{c(c+1)\dots (c+m)}\over{(2m+1)!}}
{\left( -1\right)}^m u^{2m+1} } .
\label{calG}
\end{equation}
This confirms that $g(0)=0$. The cluster distribution has the following form,
in terms of the odd function ${\cal G}$:
\begin{equation}
P_t(m) = {1\over{m^*(t)}} {{\cal G}{\left( {m\over{m^*(t)}}\right) }} ,
\label{scalingPtm}
\end{equation}
where $m^* = {\left( \gamma t\right)}^{1/2}\sim {\left( \epsilon
t\right)}^{1/2}$. It explains the scaling variable ${\left( \epsilon
t\right)}^{1/2}$ used to collapse simulation data in Fig. 11.

The conditions that ${\cal G}$ must be non-negative and normalisable are
satisfied with $c=1/2$ in Eq. (\ref{Pt0depdiff}), which leads to
\begin{equation}
{\cal G}(u) = {u\over 2}e^{-{\left( u/2\right)}^2} .
\label{calGfinal}
\end{equation}
The mean cluster mass (cluster length in the original problem) is easily
obtained as
\begin{equation}
{\langle m\rangle}_t \approx \sqrt{\pi} {\left( \gamma t\right)}^{1/2} .
\label{mdifdepfinal}
\end{equation}
Considering relation (\ref{rates}), the amplitude of cluster length scaling is
$B=\sqrt{2\pi}\epsilon^{1/2}\approx 2.507\epsilon^{1/2}$. It quantitatively
agrees with the result obtained in simulations (Sec. \ref{Simulationsdepdif}).

\section{Conclusion}
\label{secconclusion}

We studied two one-dimensional exclusion models with
particle diffusion, reversible or irreversible attachment to clusters and
deposition mechanisms.

In model I, starting from a randomly filled lattice,
only particle diffusion is allowed, with small detachment rates $\epsilon$ for
particles at the edges of the clusters. Simulation results show an initial
regime with formation of small clusters, a regime of cluster size growth as
$d\sim t^{1/3}$ and a regime of cluster size saturation at $d\sim
\epsilon^{-1/2}$. These results can be explained using heuristic scaling
arguments. The analytical treatment of the master equation with an
independent cluster approximation for joint probabilities distributions
predicts a saturation cluster size in quantitative agreement with numerical
data.

Model II generalizes model I in having also particle deposition: this is
allowed only at empty sites with one or two empty nearest neighbors.
Simulation results show continuous coarsening with a $t^{1/2}$ increase of
the average cluster size and an increase of the density with $t^{-1/2}$
corrections. These scaling forms are justified by analytical investigations
again using an independent cluster approximation, which provides good
quantitative agreement with the simulations.

We expect that the models presented above and the combination of different
methods to explain their scaling behaviors can be used to understand further
non-equilibrium systems. Of particular interest would be the extension of
theoretical methods (e. g. scaling approaches) to two-dimensional systems
such as adatom islands on surfaces or the extension of the one-dimensional
models to include other mechanisms that drive the systems to new
non-equilibrium steady states or which lead to anomalous coarsening.

\acknowledgements

FDAA Reis thanks the Department of Theoretical Physics at Oxford University,
where part of this work was done, for the hospitality, and acknowledges
support by CNPq and FINEP (brazilian agencies).

RB Stinchcombe acknowledges support
from the EPSRC under the Oxford Condensed Matter Theory Grants,
numbers GR/R83712/01 and GR/M04426.

\end{document}